\documentclass[amsmath,amssymb,epjc3]{svjour3}  
\smartqed  
\RequirePackage{graphicx}
\RequirePackage{amsmath}
\RequirePackage{amsfonts}
\RequirePackage{epstopdf}
\RequirePackage{epsf}
\RequirePackage{psfrag}
\RequirePackage{epsfig}
\RequirePackage{graphics}
\RequirePackage{epstopdf}
\newcommand{\be}{\begin{equation}}
\newcommand{\ee}{\end{equation}}
\newcommand{\bq}{\begin{eqnarray}}
\newcommand{\eq}{\end{eqnarray}}

\journalname{Eur. Phys. J. C}
\begin{document}

\title{\boldmath Subtleties in the beta function calculation of N=1 supersymmetric gauge theories}


\author{A. L. Cherchiglia\thanksref{e1,addr1,addr2}
        \and
        Marcos Sampaio\thanksref{e2,addr1}
				\and
				B. Hiller\thanksref{e3,addr3}
				\and
				A. P. Ba\^eta Scarpelli\thanksref{e4,addr4}.
}

\thankstext{e1}{e-mail: adriano@fisica.ufmg.br}
\thankstext{e2}{email: msampaio@fisica.ufmg.br}
\thankstext{e3}{email: brigitte@teor.fis.uc.pt}
\thankstext{e4}{email: scarpelli.apbs@dpf.gov.br}


\institute{Departamento de F\'{\i}sica - ICEx - Universidade Federal de Minas Gerais\\ P.O. BOX 702, 30.161-970, Belo Horizonte - MG - Brazil \label{addr1}
           \and
           Departamento de F\'{\i}sica Te\'orica y del Cosmos and CAFPE - Universidad de Granada\\E-18071, Granada - Spain \label{addr2}
           \and
					Departamento de F\'{\i}sica, Faculdade de Ci\^encias e Tecnologia, CFisUC, Universidade de Coimbra, 3004-516 Coimbra - Portugal\label{addr3}
					\and
					Setor T\'{e}cnico-Cient\'{\i}fico - Departamento de Pol\'\i cia Federal, Rua Hugo D'Antola, 95 - Lapa - S\~{a}o Paulo - Brazil\label{addr4}
           }

\date{Received: date / Accepted: date}

\maketitle

\begin{abstract}
We investigate some peculiarities in the calculation of the two-loop beta-function of $N=1$ supersymmetric models which are intimately related to the so-called ``Anomaly Puzzle''. There is an apparent paradox when the computation is performed in the framework of the covariant derivative background field method. In this formalism, it is obtained a finite two-loop effective action, although a non-null coefficient for the beta-function is achieved by means of the renormalized two-point function in the background field. We show that if the standard background field method is used, this two-point function has a divergent part which allows for the calculation of the beta-function via the renormalization constants, as usual. Therefore, we conjecture that this paradox has its origin in the covariant supergraph formalism itself, possibly being an artifact of the rescaling anomaly.
\end{abstract}

\section{Introduction}

In particle physics, symmetries have always been used as guides in order to construct theories to describe nature, an idea that culminated in the Standard Model (SM) itself. Although it has passed many experimental tests, the SM must be viewed as an effective theory since it does not incorporate all fundamental interactions. Therefore, extensions to it have been proposed, supersymmetry being one of the most appealing from the theoretical viewpoint. The reason lies on the elegance of its construction since it is a natural extension of the Poincar\`e group. Much effort have been dedicated to the subject after it was first proposed in the seventies \cite{Gervais:1971ji,Volkov:1973ix,Ramond:1971gb}.

The investigation of quantum corrections to supersymmetric models has very interesting peculiarities. For example, the choice of an adequate regularization technique is a highly nontrivial subject. A model which incorporates supersymmetry is dimension specific, the spacetime dimension playing an important role in the matching between the bosonic and fermionic degrees of freedom. This poses restrictions in the use of methods based on the analytical continuation of the spacetime dimension. Regardless the method of calculation, the quantization of supersymmetric models presented some delicate questions, such as the so called ``Anomaly Puzzle'', which can be summarized as follows.

Since the works of Piguet and Ferrara \cite{Ferrara:1974pz,Clark:1978jx,Piguet:1981mu,Piguet:1981mw}, it is known that there exist two supermultiplets, one for classically conserved supercurrents and the other a chiral supermultiplet for the scale anomalies. In the first multiplet are the classically conserved currents associated to the U(1) chiral R invariance, to supersymmetry and to translation invariance, with the last two conserved at all orders, while the R current is not. In the chiral supermultiplet are the scale anomalies associated with the R current, the trace of the supersymmetry current and the trace of the energy-momentum tensor. 

At the core of the “Anomaly Puzzle” is that an unique axial current operator only exists at tree level. At quantum level the R current in the supermultiplet is broken with a coefficient which is proportional to the beta function of the scale anomaly, which can have corrections to all orders. However, according to the Adler-Bardeen theorem \cite{Adler:1969er}, the chiral anomaly is exhausted at one-loop order in perturbation theory. Remarkably, its coefficient is given by the one-loop value of the gauge beta function. Despite these differences in the quantum realizations of the axial current, Piguet and collaborators obtained a relation which links the beta function of the scale anomaly to the nonrenormalized coefficient of the axial current anomaly \cite{Piguet: 1986,Piguet: 1988}. Since the work of Novikov, Shifman, Vainshtein e Zakharov (NSVZ) \cite{Novikov:1985rd}, which obtained an exact expression for the beta function of $N=1$ Super Yang-Mills (SYM) theory, many other works \cite{Siegel:1979wq,Mas:2002xh,Stepanyantz1,Stepanyantz2,Avdeev:1980bh,Grisaru:1980nk,Caswell:1980yi,Pimenov:2009hv,Abdalla:1986vg,Fargnoli:2010mf} followed in which different regularization methods were applied and in all cases higher order corrections for the beta functions were found.

This controversial result attracted great attention and different explanations were provided. According to Shifman and Vainshtein \cite{Shifman1}, it is necessary to distinguish between the Wilson effective action and the sum of vacuum loops in the external fields. The first renormalizes only at one-loop level whereas the second receives higher order contributions due to infrared modes. On the other hand, Arkani-Hamed and Murayama \cite{Murayama} argued that the solution to the problem can be stated in a way independent of the infrared modes using the distinction between the holomorphic gauge coupling and the canonical gauge coupling. According to the authors, the dilatation anomaly is in the same multiplet of the $U_R(1)$ anomaly and is exact at one-loop order. However, due to the anomaly, the vectorial multiplet does not possess canonical kinetic terms after the dilatation. In order to get canonical kinetic terms in the vectorial multiplet an additional change in the normalization is needed. Therefore, the anomaly coming from the modified dilatation is not in the same multiplet of the $U_R(1)$ anomaly and receives contributions beyond one-loop order. This argument was criticized in some papers, since to keep the low energy physics unchanged, it is necessary to take into account the infrared modes in the derivation of the anomaly. In this sense, it is somewhat equivalent to consider the scale anomaly or to calculate the expectation value of the Wilson effective action \cite{Mas:2002xh}. In \cite{Yonekura}, it was claimed that since the definition of the gauge coupling of $N=1$ SYM may depend on the renormalization scheme, so does the beta function. They showed that the trace anomaly is one loop exact in a certain scheme, the important point being to examine in which scheme the quantum action principle is valid. To summarize, as observed in \cite{Xing-Huang}, although the $R$-current and the stress tensor belong to the same classical supercurrent, in the quantum regime it bifurcates. It is not possible to construct an unique quantum supermultiplet which contains both the stress tensor and the $R$-current.

This discussion above also appears in a perturbative analysis. Within the supergraph approach to supersymmetric models, along with on-shell infrared divergences of Yang-Mills theory, additional off-shell infrared divergences appear which must be distinguished from ultraviolet ones before renormalization is carried out. The mixing of these two types of divergences is in the center of this debate. A consistent approach should proportionate an unambiguous distinction between the infinities involved and the arbitrary scales which are byproducts of the subtractions. In dimension-type regularizations \cite{Abbott:1984pz,Grisaru:1985tc} the two-loop correction to the $\beta$-function comes from a local evanescent operator, which would be absent in the physical spacetime dimension. So, Grisaru, Milewski and Zanon conjectured that no divergence should occur beyond one loop. This is true, as we will see, depending on the approach adopted in the calculation. However, even in the case where the divergences do not occur beyond one-loop order, this does not mean the two-loop $\beta$-function vanishes. Instead, the derivation of the renormalization group functions needs some reinterpretation, which appears to be related to scaling anomaly \cite{Kraus:2001id}. In four spacetime dimensions, Differential Renormalization was applied \cite{Mas:2002xh} in the evaluation of the two-loop $\beta$ function. It was found that the result depended on infrared modes, which play a passive role. Moreover, within differential renormalization,  it was found that the scale referring to one-loop renormalization is the one to give rise to the two-loop coefficient. Because differential renormalization delivers finite renormalized amplitudes by construction, it would be interesting to investigate how renormalization is effected within an invariant framework which both operates in the physical dimension and  displays explicitly the ultraviolet behavior in terms of the renormalization constants. In \cite{Fargnoli:2010mf}, a four dimensional regularization framework was used in the computation of the two-loop coefficient of the SYM beta-function with the use of the background field method in the covariant derivative formalism \cite{Grisaru:1984ja}. Due to the non-abelian character of SYM, the background field method is urged to be applied, since it results in a huge simplification in the number of diagrams. It was found that there is no two-loop divergence which, in a first view, could indicate the absence of higher loop corrections to the beta-function. However, the renormalized two-point Green function still depends on the renormalization scale introduced at one-loop level, allowing the computation of the two-loop coefficient for the beta function, which was shown to be non-null. It is interesting to explore such result also in view of the property that, if the $n+1$ loop coefficient of the beta function of a supersymmetric theory vanishes, then it is finite to $n$ loops \cite{Mas:2002xh,Grisaru:1985tc}.

Finally, it is interesting to discuss the dependence of the $\beta$ functions in terms of the renormalization scheme applied  \cite{Yonekura}. As discussed in \cite{Murayama}, the holomorphic beta function is one loop exact whereas the canonical is given by the NSVZ $\beta$ function. The canonical gauge coupling comes from the canonical normalization  of the holomorphic gauge superfields which is anomalous and is determined by the axial anomaly. Therefore, while the holomorphic coupling  defines the Lagrangian  (for notation see please next section)
\be
{\cal{L}}_h = \frac{1}{4 g_h^2} \int d^2\theta W^a(V_h)W^a(V_h) + h.c. 
\ee
the canonical coupling defines ${\cal{L}}_c$ by replacing $1/g_h^2$ with $1/g^2 - i \theta_{YM}/(8 \pi^2)$  as well as $V_h \rightarrow g V_c$ which however are not equivalent. Because of such a rescaling anomaly the (holomorphic and canonical) coupling constants turn out to be related by a non-local relation. 

In this context, several renormalization scheme dependence issues may arise. For example, for $N=1$  massless supersymmetric QED (SQED) regularized by High Covariant Derivatives \cite{Stepanyantz3} the NSVZ $\beta$-function is naturally obtained for renormalization group functions defined in terms of the bare coupling constant and do not depend on the renormalization prescription. However, if defined in terms of the renormalized constant, the NSVZ $\beta$-function  is only obtained in a special subtraction scheme namely the NSVZ scheme. Thus, for the exact $\beta$-function it is natural to ask in which (precise)  scheme its expression holds. For instance the $\beta$-function in the minimal subtraction scheme of dimensional reduction is not given by the NSVZ $\beta$ function beyond two loop level \cite{Jack}. In any case, it should be noticed that renormalization scheme dependence arises when comparing different approaches.

In this work, motivated by our results in \cite{Fargnoli:2010mf}, we use massless SQED as a laboratory to investigate peculiarities in the calculation of beta-functions of supersymmetric theories up to two-loop order. Particularly, we would like to understand the apparent paradox found in \cite{Fargnoli:2010mf} in the simplest context. Therefore, in the present work we will use SQED as a probe. For this purpose, we will compute the two-loop coefficient of the SQED beta-function using two different approaches: the standard background field method \cite{Abbott:1980hw} and the one based on the covariant derivative formalism \cite{Grisaru:1984ja}. We will find that in the first case there is a two-loop divergence, allowing the computation of the beta-function coefficient by standard renormalization constants. It is also possible to perform the computation using the renormalized two-point Green function, furnishing the same result as before, as expected. In the second case, we obtain the same result of \cite{Fargnoli:2010mf}: there is no two-loop divergence, even though the renormalized two-point Green function depends on a renormalization scale, furnishing a non-null value for the two-loop beta-function coefficient. Therefore, we conjecture that the paradox found in \cite{Fargnoli:2010mf} has its origin in the covariant supergraph formalism itself, possibly being an artifact of the rescaling anomaly, discussed in \cite{Shifman1,Murayama,Kraus,Kraus:2001id}. Indeed, a mechanism of corrections to the one-loop result from one-loop anomalies is described in \cite{Murayama}, through the quantum breaking of holomorphy of the coupling constant.

The outline of this paper is as follows. In section II, we present the supersymmetric QED and evaluate, in the framework of the standard background field method, the two-loop $\beta$-function, using both the renormalization constants and the renormalized two-point function in the background field. In section III, the formalism of covariant derivative background field method is applied in the calculation of the two-loop beta-function of SQED. We present in section IV our conclusions and perspectives and some results of integrals are displayed in the appendix of section V.

\section{$N=1$ SQED in the standard background field method}

In the superfield formalism, the classical action of the massless $N=1$ supersymmetric quantum electrodynamics (SQED) is given by \cite{Wess:1992cp}
\be
S=\int d^{4}x d^{2}\theta \; W^{2}+\int d^{4}x d^{4} \theta \; \bar{\Phi}_{+} e^{gV} \Phi_{+}+\int d^{4}x d^{4} \theta \; \bar{\Phi}_{-} e^{gV} \Phi_{-},
\label{action}
\ee
where $\Phi$ is a chiral field that express the matter part of the action and $V$ is a real scalar superfield that contains the gauge field $A_{\mu}$ of QED as one of its components (therefore, it is the supersymmetric generalization to the gauge field). Finally, $W$ is the supersymmetric generalization to the stress tensor of QED. In terms of the superfield $V$, one has
\be
\int d^{4}x d^{2}\theta \; W^{2}=\frac{1}{2}\int d^{4}x d^{4}\theta \; V D^{\beta} \bar{D}^{2} D_{\beta} V.
\ee

The following step would be to perform the quantization of the classical theory. However, since we want to use the background field method, we have to introduce this new field at this point. For the abelian case, we will have a linear quantum-background splitting as below \cite{Gates:1983nr}
\be
V\rightarrow V+B,
\ee
where $B$ is the background gauge field. We may now perform the quantization as usual, introducing a gauge-fixing term for the quantum gauge field $V$, as well as sources \cite{Abbott:1980hw}. The relevant fact to be noticed is that, by construction, the action will be gauge invariant in the background gauge field. This must remain valid even after renormalization. Thus, the renormalization constant for the background field $Z_{B}$ will be related to the one for the gauge coupling $Z_{g}$ as follows\footnote{We are using the definitions $g_{0}=Z_{g}g$ and $B_{0}=Z^{1/2}_{B}B$, where $g_{0},\;B_{0}$ and $g,\;B$ are bare and renormalized functions, respectively.}
\be
Z_{g}Z^{1/2}_{B}=1.
\ee
Thus, in order to obtain the beta-function of the theory, we need only to compute the two-point functions with background fields as external legs.

\begin{figure}[!h]
\begin{center}
\includegraphics[scale=0.7]{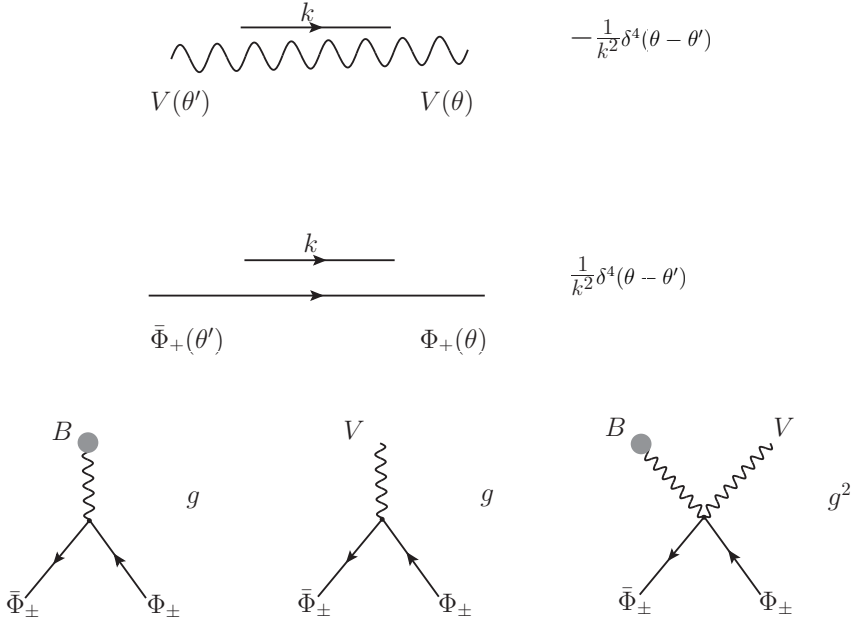}
\end{center}
\caption{Feynman rules needed for the evaluation of two point functions in the background field up to two-loop order.}
\label{feyn}
\end{figure}

As stated in the introduction, we intend to compute the two-loop corrections of the SQED beta-function. The Feynman rules can be derived from the action \cite{Gates:1983nr}, and the relevant ones for our computation are expressed in figure \ref{feyn}.

\begin{figure}[!h]
\begin{center}
\includegraphics[scale=0.7]{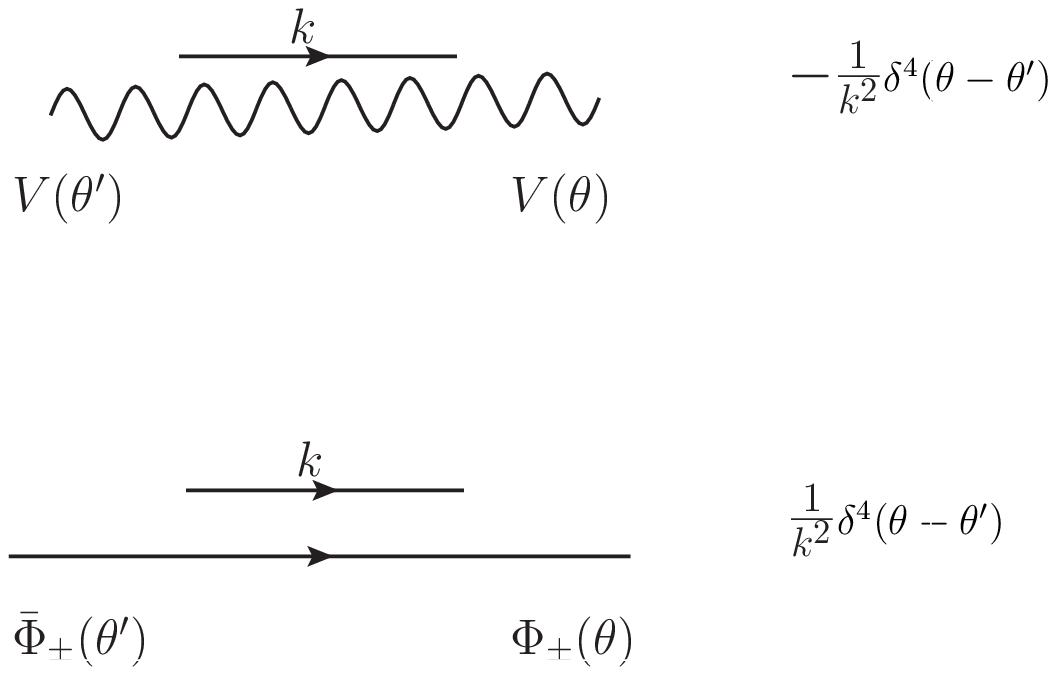}
\end{center}
\end{figure}

\begin{figure}[!h]
\begin{center}
\includegraphics[scale=0.5]{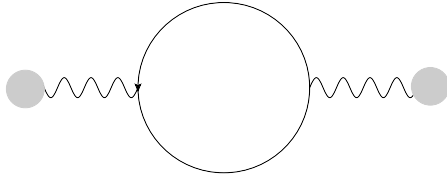}
\end{center}
\caption{1-loop diagram.}
\label{1loop-1}
\end{figure}

We start by the one-loop contribution, whose diagram\footnote{We do not include a tadpole diagram since, as we are dealing with a massless theory, it can be promptly set to zero in the Implicit Regularization formalism. However, even if such diagram was included, it would cancel out in the sum. Thus, in order to simplify the discussion, we opt to omit hereafter all the tadpole diagrams.}, using the background field method, is depicted in figure \ref{1loop-1}, furnishing the following effective action
\be
\Lambda^{(1)}\equiv2\frac{g^{2}}{2}\int_{p,\theta}B(-p,\theta)\int_{k}\left[\frac{D^{2}\bar{D}^{2}-k^{\beta\dot{\alpha}}D_{\beta}\bar{D}_{\dot{\alpha}}-k^{2}}{k^{2}(k+p)^{2}}\right]B(p,\theta),
\ee
where we have already performed the D-algebra manipulations, $\int_{k}$ stands for $\int \frac{d^{4}k}{(2\pi)^{4}}$, $\int_{p,\theta}$ for $\int \frac{d^{4}p}{(2\pi)^{4}}\int d^{4}\theta$ and $B(p,\theta)$ is the background gauge field. The factor $2$ accounts contributions from chiral fields $\Phi$ with different signs. Regarding supersymmetric definitions and conventions, we are following the ones found in \cite{Gates:1983nr}.

To proceed, we need to resort to some regularization technique. We will choose the Implicit Regularization (IReg) formalism \cite{Orimar1,Orimar2}, which, by not resorting to any kind of dimensional extension can be promptly applied in supersymmetric theories \cite{Carneiro:2003id,Ferreira:2011cv,Fargnoli:2010mf}. The method resorts recursively to the mathematical identity,
\be
\frac{1}{(p_i+k)^2-m^2}=\frac{1}{k^2-m^2}-\frac{p_i^2+2p_i\cdot k}{(k^2-m^2)\left[(p_i+ k)^2-m^2\right]},
\ee
in order to extract the external momenta, $p_i$, from the divergent integrals. In the case of massless models, a fictitious mass, $\mu^2$, is used, which is eliminated from the result by means of the limit $\mu^2 \to 0$ and of scale relations. As a byproduct, a mass scale $\lambda^2$ is introduced, which is adequate for the computation of the renormalization group functions. A detailed account on IReg can be found in \cite{Ferreira:2011cv,Cherchiglia:2010yd,Edson-massless}.

After some D-algebra manipulation the final result is
\begin{align}
\Lambda^{(1)}=&(-i)\frac{g^{2}}{2}\int_{p,\theta}B(-p,\theta)D^{\beta}\bar{D}^{2}D_{\beta}B(p,\theta)\left[I_{log}(\lambda^{2})-b\ln\left(-\frac{p^{2}}{\lambda^{2}}\right)+2b\right]+\nonumber\\
&(+i)\frac{g^{2}}{2}\int_{p,\theta}B(-p,\theta)\left[p^{\beta\dot{\alpha}}D_{\beta}\bar{D}_{\dot{\alpha}}+2k^{2}\right]\Gamma_{0}^{(1,2)}B(p,\theta),
\label{1loop}
\end{align}
with $b=i/(4\pi)^2$,
\be
I_{log}^{(l)}(\mu^2)\equiv \int_k \frac{1}{(k^2-\mu^2)^2} \ln^{l-1}\left(-\frac{k^2-\mu^2}{\lambda^2}\right)
\ee
and
\be
g^{\mu_1\cdots \mu_j}\Gamma_i^{(l,j)} \equiv \int_k \frac{\partial}{\partial k_{\mu_1}}\frac{k^{\mu_2}\cdots k^{\mu_j}}{(k^2-\mu^2)^{\frac{2+j-i}{2}}} \ln^{l-1}\left(-\frac{k^2-\mu^2}{\lambda^2}\right),
\ee
where $g^{\mu_1\cdots \mu_j}\equiv g^{\mu_1 \mu_2}\cdots g^{\mu_{j-1} \mu_j}+ \mbox{symmetric combinations}$ and the index $i$ indicates the superficial degree of divergence of the surface term. In $I_{log}(\lambda^2)$, we omitted the upper index for $l=1$.

Some comments are in order: notice that, apart from the surface term $\Gamma_{0}^{(1,2)}$, we have a gauge invariant result. This was expected, since, as we are working with the background field method, gauge invariance is explicitly maintained being broken just by regularization dependent terms. Thus we verify that the condition to preserve gauge invariance, even in supersymmetric theories, is to set surface terms to zero, as discussed in \cite{Ferreira:2011cv}. From now on, we will not display the one-loop or higher order surface terms, which will all be set to zero. Another aspect to be mentioned is the appearance of a divergent integral parametrized as a $I_{log}(\lambda^{2})$. Such term must be renormalized as usual and it will contribute to the renormalization constant $Z_{B}$, as we are going to show in the end of this section. Notice also the appearance of the renormalization scale $\lambda^2$ in the finite part as well.

\begin{figure}[!h]
\begin{center}
\includegraphics[scale=0.6]{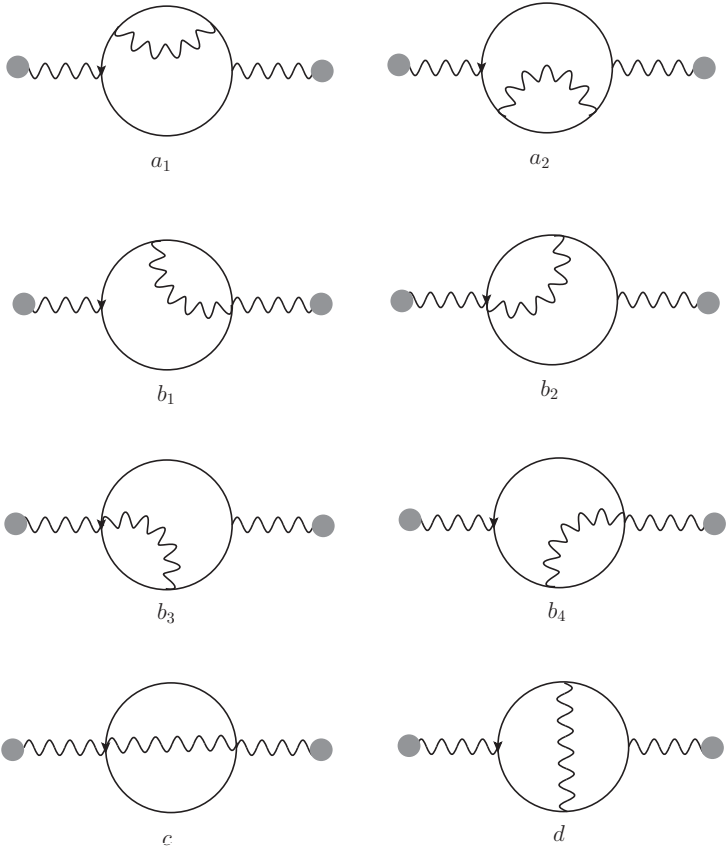}
\end{center}
\caption{2-loop diagrams.}
\label{2loop-1}
\end{figure}

We proceed now to the two-loop contributions, which are depicted in figure \ref{2loop-1}. The first diagram furnishes,
\begin{align}
\Lambda^{(2)}_{a_{1}}=&2\frac{g^{4}}{2}\int_{p,\theta}B(-p,\theta)\int_{k}\left[\frac{\bar{D}^{2}D^{2}+k^{\dot{\alpha}\beta}\bar{D}_{\dot{\alpha}}D_{\beta}-k^{2}}{k^{2}(k-p)^{2}}\right](ib)
\left[\ln\left(-\frac{p^{2}}{\lambda^{2}}\right)-2\right]B(p,\theta)\nonumber\\
&+2\frac{g^{4}}{2}\int_{p,\theta}B(-p,\theta)\int_{k}\left[\frac{\bar{D}^{2}D^{2}+k^{\dot{\alpha}\beta}\bar{D}_{\dot{\alpha}}D_{\beta}-k^{2}}{k^{2}(k-p)^{2}}\right](-i)I_{log}(\lambda^{2})B(p,\theta),
\end{align}
where the second line is a subdivergence which is be to canceled out by a one-loop counterterm according to the subtractions stated by the Bogoliubov's recursion formula. Instead of evaluating the first line, we proceed to the next diagram (the reason will be apparent shortly), which has the result quoted below,
\begin{align}
\Lambda^{(2)}_{a_{2}}=&2\frac{g^{4}}{2}\int_{p,\theta}B(-p,\theta)\int_{k}\left[\frac{D^{2}\bar{D}^{2}-k^{\beta\dot{\alpha}}D_{\beta}\bar{D}_{\dot{\alpha}}-k^{2}}{k^{2}(k+p)^{2}}\right](ib)
\left[\ln\left(-\frac{p^{2}}{\lambda^{2}}\right)-2\right]B(p,\theta)\nonumber\\
&+2\frac{g^{4}}{2}\int_{p,\theta}B(-p,\theta)\int_{k}\left[\frac{D^{2}\bar{D}^{2}-k^{\beta\dot{\alpha}}D_{\beta}\bar{D}_{\dot{\alpha}}-k^{2}}{k^{2}(k+p)^{2}}\right](-i)I_{log}(\lambda^{2})B(p,\theta).
\end{align}
Once again, the last line will be subtracted by applying Bogoliubov's recursion formula.

The reason why we have not performed the computation of the integral in $k$ will become apparent now. After performing the D-algebra on the diagrams $b_{1}\cdots b_{4}$ we see that they all can be expressed in terms of $\Lambda^{(2)}_{a_{1}}$ and $\Lambda^{(2)}_{a_{2}}$
as below
\be
\Lambda^{(2)}_{b_{1}}=\Lambda^{(2)}_{b_{2}}=-\Lambda^{(2)}_{a_{1}}, \quad \quad \Lambda^{(2)}_{b_{3}}=\Lambda^{(2)}_{b_{4}}=-\Lambda^{(2)}_{a_{2}}.
\ee

We proceed to diagram $c$, which is given by
\begin{align}
\Lambda^{(2)}_{c}=-2\frac{g^{4}}{2}\int_{p,\theta}B(-p,\theta)\int_{k,l}\frac{1}{k^{2}l^{2}(l-k-p)^2}B(p,\theta).
\end{align}

No further analysis should be taken, since the last diagram, $d$, can be written in terms of $\Lambda^{(2)}_{a_{1}}$, $\Lambda^{(2)}_{a_{2}}$, and $\Lambda^{(2)}_{c}$
\begin{align}
\Lambda^{(2)}_{c}&=\Lambda^{(2)}_{a_{1}}+\Lambda^{(2)}_{a_{2}}-\Lambda^{(2)}_{c}-2\frac{g^{4}}{2}\int_{p,\theta}\!\!\! B(-p,\theta)p^{2}\left[(I_{log}(\lambda^{2})+b)\Gamma_{0}^{(1,2)}-b\Gamma_{0}^{(2,2)}\right]\!\! B(p,\theta)\nonumber\\
&-2\frac{g^{4}}{2}\int_{p,\theta}B(-p,\theta)\left[p^{\beta\dot{\alpha}}\bar{D}_{\dot{\alpha}}D_{\beta}(-b\Gamma_{0}^{(2,2)}+2b\Gamma_{0}^{(1,2)})\right]B(p,\theta)\nonumber\\
&-2\frac{g^{4}}{2}\int_{p,\theta}B(-p,\theta)D^{\beta}\bar{D}^{2}D_{\beta}B(p,\theta)b\left[I_{log}(\lambda^{2})-b\ln\left(-\frac{p^{2}}{\lambda^{2}}\right)-\Gamma_{0}^{(1,2)}\right]\! B(p,\theta).
\end{align}

Therefore, the final two-loop correction for the effective action, after discarding the surface terms, is given by
\begin{align}
\Lambda^{(2)}=-g^{4}\int_{p,\theta}B(-p,\theta)D^{\beta}\bar{D}^{2}D_{\beta}B(p,\theta)b\left[I_{log}(\lambda^{2})-b\ln\left(-\frac{p^{2}}{\lambda^{2}}\right)\right]B(p,\theta).
\label{2loop}
\end{align}

A curious fact is that, although we are working at two-loop level,  we have a divergent result parametrized by $I_{log}(\lambda^{2})$, which is a typical basic divergent integral (BDI) of one-loop order. Our next task is to perform the renormalization of the theory, whose bare action is
\begin{align}
S_{0}=\int d^{4}x d^{4}\theta \;&\left[ \frac{1}{2} B_{0} D^{\beta} \bar{D}^{2} D_{\beta} B_{0}+ \bar{\Phi_{0}}_{+} e^{g_{0}B_{0}} {\Phi_{0}}_{+}+ \bar{\Phi_{0}}_{-} e^{g_{0}B_{0}} {\Phi_{0}}_{-}\right]\nonumber\\
&+\mbox{gauge fixing terms}+\mbox{terms on $V_{0}$}.
\end{align}

Performing a multiplicative renormalization defined by $B_{0}=Z^{1/2}_{B}B$, $g_{0}=Z_{g}g$, and ${\Phi_{0}}_{\pm}=Z_{\Phi_{\pm}}\Phi_{\pm}$, one can easily find that the counterterm for the two-point function in the background field is given by $A\equiv Z_{B}-1$. As already mentioned, in the background field method, the relation $Z_{g}Z^{1/2}_{B}=1$ holds. Therefore, the beta function can be calculated through the two-point function renormalization constant as follows
\be
\beta\equiv \lambda\frac{\partial}{\partial \lambda} g = -g\lambda\frac{\partial}{\partial \lambda} \ln Z_{g} = -g\lambda\frac{\partial}{\partial \lambda} \ln Z_{B}^{-1/2}.
\label{beta}
\ee
Supposing that both the countertem A and the $\beta$-function have expansions in the coupling constant $g$, we arrive at the expressions
\be
\beta_{1}=\frac{1}{2}\lambda\frac{\partial}{\partial \lambda} A_{1}
\ee
and
\be
\beta_{2}=A_{1}\beta_{1}+\frac{1}{2}\lambda\frac{\partial}{\partial \lambda}\left(A_{2}-\frac{A_{1}^{2}}{2}\right),
\ee
where $\beta_{i}$ is the $i$-loop coefficient of the beta function, $A_{i}$ is the $i$-loop two-point function counterterm and $\lambda$ is the renormalization group scale. Using the one- and two-loop corrections obtained in equations (\ref{1loop}) and (\ref{2loop}), respectively, and adopting a minimal subtraction scheme (which in the IReg framework amounts to the subtraction of BDI's only),
we have
\be
A_{1}=iI_{log}(\lambda^{2})\,\,\,\mbox{and}\,\,\, A_{2}=2bI_{log}(\lambda^{2}).
\ee
Finally, by using
\be
\lambda\frac{\partial}{\partial \lambda} I_{log}(\lambda^{2})=2\lambda^{2}\frac{\partial}{\partial \lambda^{2}}I_{log}(\lambda^{2}) = -2b,
\ee
with $b=\frac{i}{(4\pi)^{2}}$, we obtain the contributions for the beta function of SQED up to two-loop level in the IReg formalism,
\be
\beta=\frac{1}{(4\pi)^{2}}g^{3}+\frac{1}{8(4\pi^{2})^{2}}g^{5}+\mathcal{O}(g^{7}),
\label{betafunction}
\ee
which agrees with previous ones found in the literature \cite{Seijas:2007af,Vainshtein:1986ja,Shifman:1985fi,Novikov:1985rd}.\footnote{To obtain the results of the last three references, it should be taken into account that our definition for the coupling constant \cite{Gates:1983nr} differs from the usual one by a factor $\sqrt{2}$.}

It should be emphasized that we performed the above computation using only the renormalization constant for the background field, $Z_{B}$. Alternatively, the computation can be performed using the renormalization group equation, in which only the renormalized effective action is considered. As a consistency check, we also compute the two-loop SQED beta-function using the renormalized effective action.

As can be found, for example, in \cite{Peskin:1995ev}, the renormalization group equation reads
\be
\left[\lambda\frac{\partial}{\partial \lambda} +\beta\frac{\partial}{\partial g}-\gamma\right]G_{ren}^{(2)}(g,\lambda)=0,
\label{rg}
\ee
where
\be
\beta\equiv \lambda\frac{\partial}{\partial \lambda} g = -g\lambda\frac{\partial}{\partial \lambda} \ln Z_{g} \,\,\, \mbox{and} \,\,\, \gamma\equiv \lambda\frac{\partial}{\partial \lambda} \ln Z_{B}.
\ee
Since in the background field method  $Z_{g}Z^{1/2}_{B}=1$, we find that $\gamma=\frac{2\beta}{g}$. Thus,
\be
\left[\lambda\frac{\partial}{\partial \lambda} +\beta\left(\frac{\partial}{\partial g}-\frac{2}{g}\right)\right]G_{ren}^{(2)}(g,\lambda)=0.
\ee
From equations (\ref{1loop}) and (\ref{2loop}), we obtain the renormalized two-point function as below
\begin{align}
G_{ren}^{(2)}(g,\lambda)=\frac{1}{2}\int_{p,\theta}B(-p,\theta)D^{\beta}\bar{D}^{2}D_{\beta}B(p,\theta)\left\{1+ib\left[\ln\left(-\frac{p^{2}}{\lambda^{2}}\right)-2\right]g^{2}\right.\nonumber\\
\left.+2b^{2}\ln\left(-\frac{p^{2}}{\lambda^{2}}\right)g^{4}\right\}.
\label{Gren}
\end{align}
Replacing the expression above in equation (\ref{rg}),  we finally obtain the same expression as in equation (\ref{betafunction}).

We finish this section with some comments. In \cite{Grisaru:1985tc}, the authors conjectured that no divergence should occur beyond one-loop in $N=1$ SYM theory if the calculation is performed in the physical dimension of the model. Here, we performed the two-loop calculation of the two-point function with the use of the standard background field method for SQED and found a divergent result. It should be observed, however, that the divergence found is typical of a one-loop calculation. In the next section, we will adopt an alternative approach, which is the background field method in the covariant supergraph formalism. It is just what was done in \cite{Fargnoli:2010mf} for SYM. In this case we will see that the result will match with the conjecture of \cite{Grisaru:1985tc}, although this does not imply the two-loop beta-function is null.

\section{The SQED $\beta$-function in the covariant supergraph formalism}
\label{sec3}

We now perform the whole calculation again, by using a different approach: the background field method based on the covariant supergraph formalism \cite{Grisaru:1984ja}. The reason is the following: in \cite{Fargnoli:2010mf}, in which the aforementioned formalism was applied, the Super Yang-Mills  theory was studied and, particularly, the beta-function coefficients were computed up to two-loop order. It was found there that no two-loop divergence appeared,which could be taken as an indication of a null two-loop coefficient. However, the renormalized effective action at two-loop order still carried a dependence on the renormalization scale $\lambda$, allowing the computation of the beta function from the renormalization group equation, furnishing a non-null result for the two-loop coefficient. Therefore, it seems that there is an inconsistency, since both approaches should be equivalent. It was conjectured there that this difference should have its origin in the rescaling anomaly, as suggested by \cite{Kraus:2001id} in which the author discusses that in a framework that uses the canonical coupling (as ours), a modification of the usual multiplicative renormalization program should be necessary. Therefore we will in the following use SQED as a probe to study if the same behavior occurs in this case.

A complete description of the background field method based on the covariant supergraph formalism can be found in \cite{Grisaru:1984ja,Gates:1983nr}. The main idea is to take a step backward and work, from the beginning, with an action that depends only on background covariant derivatives. This way, all dependence on the background field will only appear implicitly. The main gain on this approach is the reduction in the number of diagrams. For instance, the two-loop correction we are going to compute requires considering only three diagrams, instead of eight, as in the previous formalism. Explicitly, the quadratic part of the action in the gauge fields we are going to work with is given by
\be
S=-\int d^{4}x d^{4}\theta \nabla^{\alpha}\bar{\nabla}^{2}\nabla_{\alpha},
\ee
where $\nabla$ is a covariant derivative in the unsplit gauge field ($V+B$). The splitting can be carried out, in the quantum-chiral but background-vector representation, as
\be
\nabla_{\alpha}=e^{-V}\boldsymbol{\nabla}_{\alpha}e^{V}, \quad \quad \bar{\nabla}_{\dot{\alpha}}=\bar{\boldsymbol{\nabla}}_{\dot{\alpha}},
\ee
being $\boldsymbol{\nabla}$ background covariant derivatives. The quantization procedure should be carried out from this point, adding gauge fixing and source terms, as usual. Chiral fields should also be included to define SQED properly (this fields must also be written in the background covariant representation). After all these considerations, we obtain the covariant Feynman rules \cite{Gates:1983nr} which, applied to our case, furnish the following one-loop effective action (the diagram depicted is the same of fig. \ref{1loop-1})
\be
\mathcal{A}^{(1)}=(-ig^{2})\int_{p,\theta}\mathbf{W}^{\alpha}(p)\mathbf{\Gamma}_{\alpha}(-p)\int_{k}\frac{1}{k^{2}}\frac{1}{(k+p)^{2}}.
\ee

Notice that our result has no explicit dependence on the background field $B$. It appears only through the field strength $\mathbf{W}^{\alpha}$ and the spinor connection $\mathbf{\Gamma}_{\alpha}$. This is a feature of the method, since background covariance, by construction, is always maintained. We obtain:
\be
\mathcal{A}^{(1)}=(-ig^{2})\int_{p,\theta}\mathbf{W}^{\alpha}(p)\mathbf{\Gamma}_{\alpha}(-p)\left[I_{log}(\lambda^{2})-b\ln\left(-\frac{p^{2}}{\lambda^{2}}\right)+2b\right].
\ee

To conclude the one-loop calculation, we write our result in terms of an explicit background gauge field. For this purpose, we recur to the definitions of $\mathbf{W}^{\alpha}$ and $\mathbf{\Gamma}_{\alpha}$ found in \cite{Abbott:1984pz},
\be
\mathbf{\Gamma}_{\alpha}=iD_{\alpha}\frac{B}{2} \,\,\, \mbox{and} \,\,\, \mathbf{W}_{\alpha}=i\bar{D}^{2}D_{\alpha}B,
\label{def}
\ee
such that our final one-loop result reads
\be
\mathcal{A}^{(1)}=(-i)\frac{g^{2}}{2}\int_{p,\theta}B(-p,\theta)D^{\beta}\bar{D}^{2}D_{\beta}B(p,\theta)\left[I_{log}(\lambda^{2})-b\ln\left(-\frac{p^{2}}{\lambda^{2}}\right)+2b\right].
\ee

Comparing with our previous expression, eq. (\ref{1loop}), we notice that there is no dependence on a surface term this time. We conjecture that this feature may be a consequence of the method which, by maintaining background covariance from the beginning, have automatically canceled all gauge-breaking terms that could occur. We also remark that this is the only difference between the two results.

\begin{figure}[!h]
\begin{center}
\includegraphics[scale=0.8]{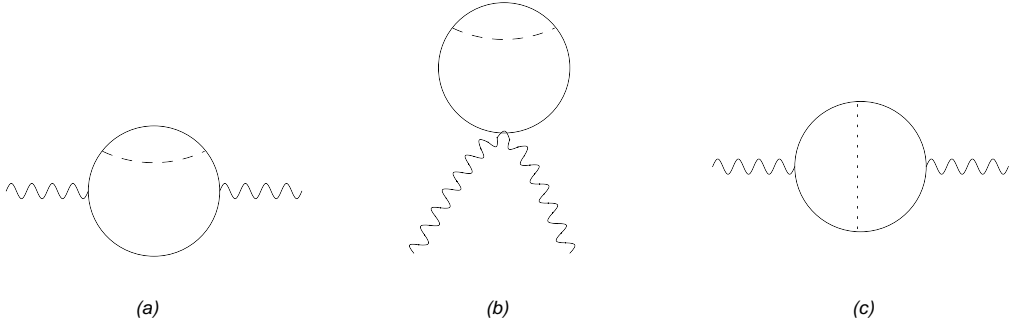}
\end{center}
\caption{2-loop diagrams in the background covariant approach.}
\label{2loop-2}
\end{figure}

We proceed now to the two-loop contribution whose diagrams are depicted in figure \ref{2loop-2}.

As in the one-loop case, the effective action will have only an implicit dependence on the background gauge field, through the field strength $\mathbf{W}^{\alpha}$ and the vector connection $\mathbf{\Gamma}^{\underline{a}}$, as the expressions for the diagrams above reveal,
\begin{align}
\mathcal{A}^{(2)}_{a}=4g^{4}\int_{p,\theta}&\left[2\mathbf{W}^{\alpha}(-p,\theta)\mathbf{\bar{W}}^{\dot{\alpha}}(p,\theta)\left(I_{1}\right)_{\alpha\dot{\alpha}}
+\frac{1}{2}\boldsymbol{\nabla}^{\alpha}\mathbf{W}_{\alpha}(-p,\theta)\boldsymbol{\nabla}^{\beta}\mathbf{W}_{\beta}(p,\theta)I_{2}\right.\nonumber\\
&\quad\left.-\frac{1}{2}\mathbf{\Gamma}^{\underline{a}}(-p,\theta)\mathbf{\Gamma}^{\underline{b}}(p,\theta)\left(I_{3}\right)_{\underline{a}\underline{b}}\right], \\
\mathcal{A}^{(2)}_{b}=4g^{4}\int_{p,\theta}&\left[\frac{1}{2}\mathbf{\Gamma}^{\underline{a}}(-p,\theta)\Gamma_{\underline{a}}(p,\theta)I_{4}\right]
\end{align}
and
\be
\mathcal{A}^{(2)}_{c}=4g^{4}\int_{p,\theta}\left[\frac{1}{4}\mathbf{\Gamma}^{\underline{a}}(-p,\theta)\mathbf{\Gamma}^{\underline{b}}(p,\theta)\left(I_{5}\right)_{\underline{a}\underline{b}}
+\frac{1}{4}\boldsymbol{\nabla}^{\alpha}\mathbf{W}_{\alpha}(-p,\theta)\boldsymbol{\nabla}^{\beta}\mathbf{W}_{\beta}(p,\theta)I_{6}\right],
\ee
where $I_{i}$ are the following integrals
\begin{align}
&\left(I_{1}\right)_{\alpha\dot{\alpha}}\equiv\sigma^{\mu}_{\alpha\dot{\alpha}}\left(I_{1}\right)_{\mu}=\sigma^{\mu}_{\alpha\dot{\alpha}}(-i)^{2}\int_{q,k}\frac{(p-k)_{\mu}}{q^{2}(q+k)^{2}k^{4}(k-p)^{2}},\\
&I_{2}\equiv(-i)^{2}\int_{q,k}\frac{1}{q^{2}(q+k)^{2}k^{4}(k-p)^{2}},\\
&\left(I_{3}\right)_{\underline{a}\underline{b}}\equiv\sigma^{\mu}_{\alpha\dot{\alpha}}\sigma^{\nu}_{\beta\dot{\beta}}\left(I_{3}\right)_{\mu\nu}=
\sigma^{\mu}_{\alpha\dot{\alpha}}\sigma^{\nu}_{\beta\dot{\beta}}(-i)^{2}\int_{q,k}\frac{4k_{\mu}k_{\nu}-2p_{\mu}k_{\nu}-2k_{\mu}p_{\nu}+p_{\mu}p_{\nu}}{q^{2}(q+k)^{2}k^{4}(k-p)^{2}},\\
&I_{4}\equiv(-i)^{2}\int_{q,k}\frac{1}{q^{2}(q+k)^{2}k^{4}},\\
&\left(I_{5}\right)_{\underline{a}\underline{b}}\equiv\sigma^{\mu}_{\alpha\dot{\alpha}}\sigma^{\nu}_{\beta\dot{\beta}}\left(I_{5}\right)_{\mu\nu}=
\sigma^{\mu}_{\alpha\dot{\alpha}}\sigma^{\nu}_{\beta\dot{\beta}}(-i)^{2}\int_{q,k}\frac{4k_{\mu}q_{\nu}-2k_{\mu}p_{\nu}+2p_{\mu}q_{\nu}-p_{\mu}p_{\nu}}{q^{2}(q+k)^{2}k^{2}(k+p)^{2}(q-p)^{2}},\\
&I_{6}\equiv(-i)^{2}\int_{q,k}\frac{1}{q^{2}(q+k)^{2}k^{2}(k+p)^{2}(q-p)^{2}}.
\end{align}

Since now we have some integrals with off-shell infrared divergences, we will explain with same detail the treatment of the first integral. Within the IReg procedures, we have
\begin{align}
\left(I_{1}\right)_{\mu}=(-i)^{2}\int_{k}\frac{(p-k)_{\mu}}{(k^{2}-\mu^{2})^{2}[(k-p)^{2}-\mu^{2}]}\left[I_{log}(\lambda^{2})-b\ln\left(-\frac{(k^{2}-\mu^{2})}{\lambda^{2}}\right)+2b\right],
\end{align}
where we have to notice the inclusion of the fictitious mass $\mu^{2}$, which must be added in order to regularize the infrared divergence. There is also an UV divergence parametrized as a $I_{log}(\lambda^{2})$ which is just an one-loop subdivergence that is going to be canceled by the application of Bogoliubov's recursion formula. After the subtraction of the subdivergence one obtains
\begin{align}
\left(I_{1}\right)_{\mu}=(-i)^{2}\left[p_{\mu}F+bU_{\mu}^{(2)}-2bU_{\mu}\right],
\end{align}
with
\begin{align}
U_{\mu} &\equiv \int_{k}\frac{k_{\mu}}{k^4(k-p)^{2}},\\
U_{\mu}^{(2)}&\equiv\int_{k}\frac{k_{\mu}}{k^{4}(k-p)^{2}}\ln\left(-\frac{k^{2}}{\lambda^{2}}\right)
\end{align}
and
\be
F\equiv\lim_{\mu^{2}\rightarrow 0}\int_{k}\frac{1}{(k^{2}-\mu^{2})^{2}[(k-p)^{2}-\mu^{2}]}\left[-b\ln\left(-\frac{(k^{2}-\mu^{2})}{\lambda^{2}}\right)+2b\right].
\ee
As we can see, the infrared divergence is concentrated in the $F$ integral. To proper treat the IR divergence, one could, for example, resort to the IReg generalization presented in \cite{Fargnoli:2010ec}. However, in the present case the integral $F$ will cancel out with other contributions, not requiring any further treatment. For the other integrals, we obtain
\begin{align}
I_{2}=&(-i)^{2}F,\\
\left(I_{3}\right)_{\mu\nu}=&(-i)^{2}\left[p_{\mu}p_{\nu}F+2b (p_{\mu}U_{\nu}^{(2)}-2 p_{\mu}U_{\nu}+ p_{\nu}U_{\mu}^{(2)}-2 p_{\nu}U_{\mu}-2 U_{\mu\nu}^{(2)}+4 U_{\mu\nu})\right],\\
I_{4}=&(-i)^{2}\left[-bI_{log}^{(2)}(\lambda^{2})+2bI_{log}(\lambda^{2})+\frac{b^{2}}{2}\ln^{2}\left(-\frac{p^{2}}{\lambda^{2}}\right)-b^{2}\ln\left(-\frac{p^{2}}{\lambda^{2}}\right)+p^{2}F\right], \\
\left(I_{5}\right)_{\mu\nu}=&(-i)^{2}\left[4I_{\mu\nu}^{\mathcal{O}_{2}}-2p_{\nu}I_{\mu}^{\mathcal{O}} +2p_{\mu}\bar{I}_{\nu}^{\mathcal{O}} -p_{\mu}p_{\nu}I^{\mathcal{O}}\right],\\
I_{6}=&I^{\mathcal{O}},
\end{align}
where we defined
\begin{align}
U_{\mu\nu} &\equiv \int_{k}\frac{k_{\mu}k_{\nu}}{k^{4}(k-p)^{2}},\\
U_{\mu\nu}^{(2)}&\equiv\int_{k}\frac{k_{\mu}k_{\nu}}{k^{4}(k-p)^{2}}\ln\left(-\frac{k^{2}}{\lambda^{2}}\right),\\
I^{\mathcal{O}}&\equiv\int_{q,k}\frac{1}{q^{2}(q+k)^{2}k^{2}(k+p)^{2}(q-p)^{2}},\\
I_{\mu}^{\mathcal{O}}&\equiv\int_{q,k}\frac{k_{\mu}}{q^{2}(q+k)^{2}k^{2}(k+p)^{2}(q-p)^{2}},\\	
\bar{I}_{\nu}^{\mathcal{O}}&\equiv\int_{q,k}\frac{q_{\nu}}{q^{2}(q+k)^{2}k^{2}(k+p)^{2}(q-p)^{2}},\\
I_{\mu\nu}^{\mathcal{O}_{2}}&\equiv\int_{q,k}\frac{k_{\mu}q_{\nu}}{q^{2}(q+k)^{2}k^{2}(k+p)^{2}(q-p)^{2}}.
\end{align}

The results of the integrals can be found in the appendix. We now proceed noticing that, in the effective actions $\mathcal{A}^{(2)}_{i}$, we have different structures in terms of the field strength and vector connection. For reasons that are going to be apparent soon, we choose to group all the contributions proportional to the vector connection, obtaining
\begin{align}
\mathcal{A}^{(2)}_{\Gamma}=4g^{4}\int_{p,\theta}\mathbf{\Gamma}^{\underline{a}}(-p,\theta)\mathbf{\Gamma}^{\underline{b}}(p,\theta)
\left[-\frac{1}{2}\left(I_{3}\right)_{\underline{a}\underline{b}}+\frac{1}{2}\left(g_{\underline{a}\underline{b}}I_{4}\right)+\frac{1}{4}\left(I_{5}\right)_{\underline{a}\underline{b}}\right].
\end{align}

Replacing the values of the integrals found in the appendix, we have, after discarding the surface terms
\begin{align}
\mathcal{A}^{(2)}_{\Gamma}=4g^{4}\int_{p,\theta}\mathbf{\Gamma}^{\underline{a}}(-p,\theta)\mathbf{\Gamma}^{\underline{b}}(p,\theta)
\left(\frac{p_{\underline{a}}p_{\underline{b}}}{p^{2}}-g_{\underline{a}\underline{b}}\right)&\left[b^2\ln\left(-\frac{p^{2}}{\lambda^{2}}\right)
+\frac{b^2\pi^{2}}{36}\right.\nonumber\\&\left.+\frac{b^{2}\zeta(3)}{2}-\frac{8b^{2}}{3}+\frac{Fp^{2}}{2}\right].
\end{align}

Notice that, although we have dealt with ultraviolet and infrared divergent integrals, the net result is finite and gauge invariant, and obeys the following relation
\begin{align}
\int d^{4}\theta\mathbf{\Gamma}^{\underline{a}}(-p,\theta)\mathbf{\Gamma}^{\underline{b}}(p,\theta)\left(\frac{p_{\underline{a}}p_{\underline{b}}}{p^{2}}-g_{\underline{a}\underline{b}}\right)=\frac{3}{2}\int d^{2}\theta \mathbf{W}^{\alpha}\mathbf{W}_{\alpha}.
\end{align}
Since we also have the relations,
\be
\int d^{4}\theta \mathbf{W}^{\alpha}(-p,\theta)p_{\alpha\dot{\alpha}}\mathbf{\bar{W}}^{\dot{\alpha}}(p,\theta)=\frac{1}{2}\int d^{2}\theta \mathbf{W}^{\alpha}(-p,\theta)\mathbf{W}_{\alpha}(p,\theta)
\ee
and
\be
\int d^{4}\theta \boldsymbol{\nabla}^{\alpha} \mathbf{W}_{\alpha}(-p,\theta)\boldsymbol{\nabla}^{\beta}\mathbf{W}_{\beta}(p,\theta)=-\frac{1}{2}\int d^{2}\theta \mathbf{W}^{\alpha}(-p,\theta)\mathbf{W}_{\alpha}(p,\theta),
\ee
and $\left(I_{1}\right)_{\alpha\dot{\alpha}} \propto p_{\alpha\dot{\alpha}}$, we finally obtain the two-loop effective action,
\be
\mathcal{A}^{(2)}=4g^4\int_{p} d^{2}\theta \mathbf{W}^{\alpha}(-p,\theta)\mathbf{W}_{\alpha}(p,\theta)b^{2}\left[\frac{1}{2}\ln\left(-\frac{p^{2}}{\lambda^{2}}\right)+\frac{\pi^{2}}{24}+\frac{3\zeta(3)}{2}-2\right],
\ee
which, written in terms of the background field by means of eq. (\ref{def}), is given by
\be
\mathcal{A}^{(2)}=\frac{g^4}{2}\int_{p,\theta}B(-p,\theta)D^{\beta}\bar{D}^{2}D_{\beta}B(p,\theta)b^{2}\left[2\ln\left(-\frac{p^{2}}{\lambda^{2}}\right)+\frac{\pi^{2}}{6}+6\zeta(3)-8\right].
\ee

Some comments are in order. One should compare the result above with the one expressed by eq. (\ref{2loop}) which was obtained using the standard background field method. A notorious difference is the disappearance of the UV divergent integral. This means that there isn't a two-loop contribution for the renormalization constant $Z_{B}$, which, at a first view, could lead one to think that the two-loop coefficient of the SQED beta-function is null. However, one could also compute the beta-function using the renormalization group equation. For this purpose, the following renormalized two-point function is needed
\begin{align}
G_{ren}^{(2)}(g,\lambda)=\frac{1}{2}\int_{p,\theta}B(-p,\theta)D^{\beta}\bar{D}^{2}D_{\beta}B(p,\theta)\left\{1+ib\left[\ln\left(-\frac{p^{2}}{\lambda^{2}}\right)-2\right]g^{2}\right.\nonumber\\
\left.+2b^{2}\left[\ln\left(-\frac{p^{2}}{\lambda^{2}}\right)+\frac{\pi^{2}}{12}+3\zeta(3)-4\right]g^{4}\right\}.
\end{align}
By comparison with eq. (\ref{Gren}), one notices that in both methods (standard and covariant derivative background field method) the dependence of the renormalized two-point function in the renormalization scale $\lambda$ is the same. Since this is the only relevant part for the computation of the SQED beta-function, we find that in both cases there is a non-null two-loop beta-function coefficient
given by
\be
\beta_{2}=\frac{1}{8(4\pi^{2})^{2}}g^{5}.
\ee

\section{Discussion of the results and perspectives}

In this paper we have studied massless SQED up to two-loop order. Our purpose was to study the intriguing fact that different approaches in the use of the background field method result in the existence or not of a divergent part in two-loop calculations in $N=1$ supersymmetric theories, even though the corresponding $\beta$-function coefficient is the same. We used the Implicit Regularization framework, since it operates in the physical dimension of the theory (respecting supersymmetry) as well as displays in a clear way UV and IR divergences and regularization dependent surface terms. The use of the background field method simplifies considerably the calculations by reducing the computation of the beta-function to the knowledge of two-point functions in the background field. The two approaches used were the following: the standard \cite{Abbott:1980hw} and the covariant derivative \cite{Grisaru:1984ja} background field method. In the first case, we obtained that the one and two-loop effective action contained a divergence. Therefore, the beta-function could be computed in the usual way, by defining a renormalization constant in the background field. On the other hand, by using the covariant derivative background field method, we obtained that the two-loop effective action had no divergence. This could imply that the beta-function would not receive higher order corrections. However, the renormalized two-point function still depended on the mass scale $\lambda$, which allowed us to obtain the two-loop $\beta$-function coefficient. Both approaches yielded the same result
\be
\beta=\frac{1}{(4\pi)^{2}}g^{3}+\frac{1}{8(4\pi^{2})^{2}}g^{5}+\mathcal{O}(g^{7}),
\ee
coinciding with the one obtained before in the literature \cite{Novikov:1985rd,Seijas:2007af,Vainshtein:1986ja,Shifman:1985fi}. It should be noticed that even in the case of the standard background method  the beta-function could be computed by using the renormalization group equation, which delivered the same result as before.

We emphasize that our main point is to find out if the computation via the standard or covariant derivative background field method could, respectively, give rise or not to the explicit divergent behavior at two loop order, yet there is no doubt about the value of the two loop correction to the beta function as they agree (computing via the renormalization constants or via RG equation). In other words, there is no doubt about renormalization scheme in our analysis, since the value of the beta function obtained in both methods coincide with each other corroborating the universality of the two loop coefficients of the $\beta$-function.  The question would be in which method the multiplicative renormalization program is still applicable, since the divergent behavior of the effective action in both methods is not the same.

This particularity for the calculation with the covariant derivative background field method was already obtained in the context of SYM theory \cite{Fargnoli:2010mf}. Therefore, we found out with our computation that the above behavior is not characteristic of the SYM theory, being shared by the SQED theory as well. This allows us to conjecture that the reason may lie on the rescaling anomaly and the usual multiplicative renormalization program should be modified as suggested by \cite{Kraus:2001id}, being inherent to the definition of the covariant derivative background field method itself.

As perspectives we should include the study of how exactly the rescaling anomaly manifests itself in the covariant derivative background field method. Thus, one expects to be able to introduce some modifications in the usual multiplicative renormalization in order to solve this controversy in the computation of the beta function.

\appendix

\section{List of integrals used in this work}

For the integrals needed in section \ref{sec3}, we have the following results:
\begin{align}
U_{\mu} &\equiv \int_{k}\frac{k_{\mu}}{k^4(k-p)^{2}}=b \frac{p_\mu}{p^2},\\
U_{\mu}^{(2)}&\equiv\int_{k}\frac{k_{\mu}}{k^{4}(k-p)^{2}}\ln\left(-\frac{k^{2}}{\lambda^{2}}\right)=b \frac{p_\mu}{p^2}\ln \left(-\frac{p^2}{\lambda^2}\right), \\
U_{\mu\nu} &\equiv \int_{k}\frac{k_{\mu}k_{\nu}}{k^{4}(k-p)^{2}}=\frac{g_{\mu \nu}}{4}\left\{I_{log}(\lambda^2) -b \ln \left(-\frac{p^2}{\lambda^2}\right) +2b \right\}  + \frac b2 \frac{p_\mu p_\nu}{p^2},\\
U_{\mu\nu}^{(2)}&\equiv\int_{k}\frac{k_{\mu}k_{\nu}}{k^{4}(k-p)^{2}}\ln\left(-\frac{k^{2}}{\lambda^{2}}\right) =
\frac{g_{\mu \nu}}{8}\left\{2 I_{log}^{(2)}(\lambda^2)+I_{log}(\lambda^2) -b \ln^2 \left(-\frac{p^2}{\lambda^2}\right) \right.\nonumber \\
&\left. + b \ln \left(-\frac{p^2}{\lambda^2}\right) + b \right\} + \frac{p_\mu p_\nu}{p^2}\left\{ \frac b4 +\frac b2 \ln \left(-\frac{p^2}{\lambda^2}\right) \right\},\\
I^{\mathcal{O}}&\equiv\int_{q,k}\frac{1}{q^{2}(q+k)^{2}k^{2}(k+p)^{2}(q-p)^{2}}= \frac{6 \zeta(3)b^2}{p^2} ,\\
I_{\mu}^{\mathcal{O}}&\equiv\int_{q,k}\frac{k_{\mu}}{q^{2}(q+k)^{2}k^{2}(k+p)^{2}(q-p)^{2}}=-\frac{p_\mu}{2}I^{\mathcal{O}},\\	
\bar{I}_{\nu}^{\mathcal{O}}&\equiv\int_{q,k}\frac{q_{\nu}}{q^{2}(q+k)^{2}k^{2}(k+p)^{2}(q-p)^{2}}= \frac{p_\mu}{2}I^{\mathcal{O}},\\
I_{\mu\nu}^{\mathcal{O}_{2}}&\equiv\int_{q,k}\frac{k_{\mu}q_{\nu}}{q^{2}(q+k)^{2}k^{2}(k+p)^{2}(q-p)^{2}}= -g_{\mu \nu}\left\{ \frac b4 I_{log}(\lambda^2)
-\frac{b^2}{4} \ln \left(-\frac{p^2}{\lambda^2}\right) \right. \nonumber \\
&\left. -\frac{p^2}{12} I^{\mathcal{O}}+\frac{11}{12}b^2 - \frac{\pi^2}{36}b^2\right\} -\frac{p_\mu p_\nu}{p^2}\left\{ \frac{p^2}{3}I^{\mathcal{O}}-\frac 16 b^2 +\frac{\pi^2}{36} b^2 \right\}.
\end{align}

\begin{acknowledgements}

A. Cherchiglia acknowledges fruitful discussions with M. Perez-Victoria and thanks Universidad de Granada for the kind hospitality.  
A. Cherchiglia and M. Sampaio acknowledge financial support by FAPEMIG and CNPq, Conselho Nacional de Desenvolvimento Cient\'{\i}fico e Tecnol\'{o}gico - Brazil. 

\end{acknowledgements}

\end{document}